\documentstyle[12pt]{article}

\begin{document}

\newcommand{\APSREF}{0}
\newcommand{\LETTER}{0}
%
%
\def\daga#1{{#1\mkern -9.0mu /}}
\newcommand{\Eq}[1]{Eq.~(\ref{#1})}
\newcommand{\Ref}[1]{Ref.~\cite{#1}}
\newcommand{\beq}[1]{\begin{equation}\label{#1}}
\newcommand{\eeq}{\end{equation}}
\newcommand{\bdm}{\begin{displaymath}}
\newcommand{\edm}{\end{displaymath}}
\newcommand{\beqa}[1]{\begin{eqnarray}\label{#1}}
\newcommand{\eeqa}{\end{eqnarray}}
\newcommand{\bdma}{\begin{eqnarray*}}
\newcommand{\edma}{\end{eqnarray*}}

\ifnum\APSREF=1
\newcommand{\prd}[3]{Phys. Rev. D{\bf #1}, #2 (#3)}
\newcommand{\physrep}[3]{Phys. Rep. {\bf #1}, #2 (#3)}
\newcommand{\plb}[3]{Phys. Lett. {\bf B#1}, #2 (#3)}
\newcommand{\npb}[3]{Nucl. Phys. {\bf B#1}, #2 (#3)}
\newcommand{\np}[3]{Nucl. Phys. {\bf #1}, #2 (#3)}
\newcommand{\prl}[3]{Phys. Rev. Lett. {\bf #1}, #2 (#3)}
\newcommand{\rmp}[3]{Rev. Mod. Phys. {\bf #1}, #2 (#3)}
\newcommand{\ibid}[3]{{\em ibid.} {\bf #1}, #2 (#3)}
\newcommand{\astropj}[3]{Ap. J. {\bf #1}, #2 (#3)}
\newcommand{\annphys}[3]{Ann. Phys. {\bf #1}, #2 (#3)}
\newcommand{\repprp}[3]{Rep. Prog. Phys. {\bf #1}, #2 (#3)}
\newcommand{\jmp}[3]{J. Math. Phys.  {\bf #1}, #2 (#3)}
\newcommand{\zphysc}[3]{Z. Phys. C  {\bf #1}, #2 (#3)}
\newcommand{\ijmpa}[3]{Int. J. Mod. Phys. A  {\bf #1}, #2 (#3)}
\newcommand{\mpla}[3]{Mod. Phys. Lett. A{\bf #1}, #2 (#3)}
\newcommand{\jmpa}[3]{J.  Phys A: Math. Gen.  {\bf #1}, #2 (#3)}

\else
\newcommand{\prd}[3]{Phys. Rev. D{\bf #1} (#3) #2 }
\newcommand{\physrep}[3]{Phys. Rep. {\bf #1} (#3) #2 }
\newcommand{\plb}[3]{Phys. Lett. {\bf B#1} (#3) #2 }
\newcommand{\npb}[3]{Nucl. Phys. {\bf B#1} (#3) #2 }
\newcommand{\np}[3]{Nucl. Phys. {\bf #1} (#3) #2 }
\newcommand{\prl}[3]{Phys. Rev. Lett. {\bf #1} (#3) #2 }
\newcommand{\rmp}[3]{Rev. Mod. Phys. {\bf #1} (#3) #2 }
\newcommand{\ibid}[3]{{\em ibid.} {\bf #1} (#3) #2 }
\newcommand{\astropj}[3]{Ap. J. {\bf #1} (#3) #2 }
\newcommand{\annphys}[3]{Ann. Phys. {\bf #1} (#3) #2 }
\newcommand{\repprp}[3]{Rep. Prog. Phys.  {\bf #1} (#3) #2 }
\newcommand{\jmp}[3]{J. Math. Phys.   {\bf #1} (#3) #2 }
\newcommand{\zphysc}[3]{Z. Phys. C   {\bf #1} (#3) #2 }
\newcommand{\ijmpa}[3]{Int. J. Mod. Phys. A  {\bf #1} (#3) #2 }
\newcommand{\mpla}[3]{Mod. Phys. Lett. A{\bf #1} (#3) #2 }
\newcommand{\jmpa}[3]{J.  Phys A: Math. Gen.   {\bf #1} (#3) #2 }
\fi

\ifnum\LETTER=0
\newcommand{\newsection}[1]{\section{#1}\setcounter{equation}{0}}
\fi

\newcommand{\aslash}[1]{{\rlap/#1}}
\newcommand{\splash}[1]{{#1\mkern -9.0mu /}}
\newcommand{\splashh}[2]{{#2\mkern -#1mu /}}

\baselineskip=20pt

%
%

\begin{flushright}
Preprint IFUNAM:\\
FT94-66 Dec/94.\\
hep-ph/9412256
\end{flushright}

\vskip1.5cm

\begin{center}{\LARGE\bf
Parity  symmetry restoration  in $QED_3$ at high temperature and
finite
density.\\ \vskip0.2in
}\end{center}
\begin{center}
by

{\bf{Martin Klein-Kreisler\footnote{e-mail:
mklein@teorica0.ifisicacu.unam.mx}
 and   Manuel Torres\footnote{e-mail:
manuel@teorica1.ifisicacu.unam.mx}
}}\\\vskip0.3in
{\small\it Instituto de Fisica,  UNAM,
Apdo. Postal 20-364, \\
\vspace{-2mm}
01000  Mexico, D.F., Mexico.}

\vskip1.0in
\noindent
{\bf{Abstract}}
\end{center}
\vskip0.05in
\setlength{\baselineskip}{0.2in}
We investigate the properties  of quantum electrodynamics $(QED_3)$
in two
spatial
dimensions at finite temperature and density.  The static  as well as
the
dynamical properties of the  planar plasma are calculated  using the
real time
formalism of quantum field theory.  It is shown, that due to the
presence of
the parity breaking Chern-Simons  term,
 the propagating modes  of the photon  consists of two longielliptic
waves with different values for the masses.  However,
 it is shown that  both in the high temperature
and high density limit the  parity  symmetry is restored; in this
limit the
propagating
modes reduce to the longitudinal and transverse ones, both with the
same mass.

\vskip1cm
\begin{center}
\noindent
\rule[.1in]{3.0in}{0.002in}
\end{center}
\vskip1cm
keywords: Chern-Simons, parity  breaking, symmetry restoration.

\newpage
\baselineskip=20pt

%

Lower-dimensional field theories with Chern-Simons terms have become
subjects
of
increasing interest. They possess unusual theoretical properties
\cite{jackiwplano}
and have been   proposed to  describe the physics  of   planar
systems like the
fractional quantum Hall effect  and high temperature
superconductivity
\cite{hall}.  In the case of  gauge
theories  in $2 +1$ dimensions  with massive fermions it is well
known that a
Chern-Simons parity breaking term  is induced
through    radiative corrections \cite{deser,pisarski}.
On the other hand, breaking  and restoration  of various symmetries
at high
energy and/or finite temperature has been known for some time in
quantum field
theories \cite{linde}.

In this letter we study the  plasma properties in $QED_3$, for a
one-fermion
species, including the effect of the parity breaking Chern-Simons
(C-S)  term.
In previous work  \cite{chernT} the induced C-S term  at finite
temperature
has been  calculated
by considering  a  ``zero  momentum''  approximation $ k_\mu \to 0 $;
with  $
k_\mu $ the photon
three-momentum.  More recently,  Aitchison $et\,al.$ \cite{aitchison}
analysed,
 within the imaginary time-formalism,  the range of validity
of these results  and concluded  that  this ``zero momentum''
approximation is
implicitly a low temperature approximation. We are interested in
assessing the
effect of the C-S term on the
propagating modes of  the plasma, consequently we must go beyond the
$ k_\mu
\to 0 $
limit.  Furthermore, in order to investigate  the possible parity
restoration,
we need
to consider the  high temperature limit.  In this work we find  an
exact
expression for the C-S term
that can be used as starting point to study  its  momentum dependence
as well
as
the high-temperature and high-density  regimes in
a consistent way.   We show how the  parity symmetry broken by the
C-S term is
restored
in these limits.  We also analyse the effect this term has on the
dispersion
relations of the plasma.

A formal approach to calculate the nonzero temperature  contributions
to the
photon  self-energy  is  provided  by thermal field theory (TFT).
Within TFT,
two classes of formalisms can be distinguished: the  imaginary-time
formalism
(ITF) and the real-time formalism (RTF) \cite{review}.  We found
convenient to
work within  the RTF;
the reason being that this formalism provides us directly with  the
time-ordered  Feynman
propagators \cite{kobes},  thus avoiding the problem of  continuation
from
imaginary to real energies. A detailed derivation of
real-time Feynman  rules at finite temperature and density  using
path integral methods can be found in \cite{niemi},  whereas a
canonical
approach
to  the  RTF  can be found in \cite{nieves}. It is well-known that
the RTF
requires a doubling of the degrees of freedom, so the thermal
propagator has
then a $2\!\times\!2$ matrix structure.
However, at the one-loop level one only requires the $1-1$ component
of the
propagator.
Thus,  the  fermion   propagator  at finite temperature $T$ and
chemical
potential $\mu$     is  given by
\begin{equation}\label{profer}
 S (p)   =  \left( \daga{p}  + m \right)  \left[
		{1 \over p^2 - m^2 + i \epsilon  }+ 2 \pi i
\delta(p^2 - m^2)
		\left(  \theta(p \cdot u) n_F(x)  +   \theta( - p
\cdot u)  n_F(-x)
\right)\right] ,
\end{equation}
where  $m$  and   $p_\mu$  are   the  fermion mass and momentum,
respectively.
The theory is formulated covariantly with the introduction of  the
velocity
three-vector of the medium $u_\mu$.  In \Eq{profer}, $\theta$ is the
step
function and
 $n_F(x) = 1/ \left( e^x + 1 \right)$ is the  Fermi-Dirac
distribution
with   \\ $x =  \left( p \cdot u - \mu \right)/T$.

We now take up the description of  the  photon polarization tensor in
$2+1$
dim.
It is well known that in $3+1$ dimensions  the photon polarization
tensor is
symmetric and can be decomposed
into transverse and longitudinal  modes.  However, in three
dimensional
spacetime   the  radiative corrections generate  an  anti-symmetric
parity
violation term \cite{deser}, the Chern-Simons term. Thus, the
tensorial
structure of the   photon polarization tensor  can be written as
\begin{equation}\label{pimunu}
 \Pi_{\mu\nu} (k) \, = \, \Pi_T  P_{\mu\nu}   +   \Pi_L  Q_{\mu\nu}
+
i  \tilde{\Pi}  \epsilon_{\mu\nu \alpha} k^\alpha  \,  .
\end{equation}
Here, the functions  $ \Pi_T$,  $ \Pi_L$ and   $\tilde{\Pi}$  can
depend on two
Lorentz scalars, that are chosen as :  $ \omega   =   k \cdot u $ and
$\kappa    =  \sqrt{ \left(  k \cdot u \right)^2 - k^2 }$.
The expressions and properties of  the projectors  $P_{\mu\nu}$ and
$Q_{\mu\nu}$ can be found in  reference \cite{weldon}.

The full propagator $(\Delta_{\mu\nu})$  of  the gauge boson  is
obtained
from
the vacuum  propagator    by summing all the  self-energy
polarization
insertions, which yields
\begin{equation}\label{progauge}
 \Delta_{\mu\nu} =
{ \left[  (k^2 - \Pi_L) P_{\mu\nu}  +  (k^2 - \Pi_T)  Q_{\mu\nu}
+  i   \tilde{\Pi} \epsilon_{\mu\nu\alpha} k^\alpha \right]  \over
 (k^2 - \Pi_L)   (k^2 - \Pi_T)  -  k^2  \tilde{\Pi}^2  }
+    \left(\lambda  - 1 \right) { k_\mu k_\nu \over k^4 }   \,  .
\end{equation}
where $\lambda$ is the gauge parameter. One may notice that the gauge
dependent
part is not affected by the resummation. The poles of this resummed
propagator
define  the   gauge boson propagating modes of the  system  according
to the
relation
\begin{equation}\label{polo}
 (k^2 - \Pi_L)   (k^2 - \Pi_T)  -  k^2  \tilde{\Pi}^2 = 0 \, .
\end{equation}

 Notice that  $\Delta_{\mu\nu}$    reduces to the  well known
expression
\cite{weldon}
in   the particular  limit $\tilde{\Pi} = 0$;  if that is the case,
the modes
separate into
a longitudinal and a transverse  component.
Otherwise,  the presence of the parity breaking term
mixes the longitudinal and transverse modes, so that  the normal
propagating
modes correspond  to two longielliptic waves, each with a different
value for
the mass. In fact, corresponding to the boson  propagator
\Eq{progauge} there
is an effective Maxwell equation
$ \left\{ k^2 g_{\mu\nu} - k_\mu k_\nu (1 - \lambda)  - \Pi_{\mu\nu}
\right\}
A^\nu =\!0$,
that determines the gauge field $A^\mu$.  A solution exist only when
$\omega$
and $\kappa$
are such that the matrix in curly brackets has zero determinant; this
condition
is the same as that obtained from  \Eq{polo}.  From the  plane wave
solution
for  $A^\mu $  and assuming that the wave propagates along the
$x-$axis
  we   find that the  the electric field  can be written as
$\vec{E} \propto ( 1, i \omega  \tilde{\Pi}/\left( k^2 - \Pi_T
\right)) $ or
$\vec{E }\propto ( - i  k^2  \tilde{\Pi}/ \omega \left( k^2 - \Pi_L
\right), 1
) $.  Therefore, the waves are  neither transverse nor longitudinal;
instead
the solutions correspond to  a right-handed  and left-handed
longielliptic
waves.
These modes are  described, in general,   by  complicated
dispersion
relations.
Nonetheless, we  shall be able to  explicitly  display some    of
their
properties
in  some particular   limits.

The one-loop contribution to the photon polarization tensor,
including thermal
corrections,  is given by
\begin{equation}\label{pimunu2}
 \Pi_{\mu\nu} (k) = i e^2 \int{ {d^3p  \over (2\pi)^3}
 Tr \left[  \gamma_\mu S(k + p) \gamma_\nu S(p)     \right]}  \,  ,
\end{equation}
where the fermion propagator is given by \Eq{profer}.   Because of
Lorentz
covariance  and current conservation this  tensor is of the form
given in
\Eq{pimunu}. The two   functions  $ \Pi_L$ and  $ \Pi_T$  are
obtained by
contraction with $\Pi_{\mu\nu}$, namely:
$ \Pi_L   =    (-k^2 / \kappa ^2) u^\mu u^\nu  \Pi_{\mu\nu}$ and
$ \Pi_T  =   g^{\mu \nu}  \Pi_{\mu\nu}  - \Pi_L$.
Whereas,    $\tilde{\Pi}$ can be directly read off  from the
coefficient  of
the parity-odd  term, which appears because in $2+1 $ dimensions
the trace of three Dirac matrices does not vanish:
 $Tr\left[ \gamma_\mu \gamma_\nu \gamma_\alpha \right]  =
- 2 i  \epsilon_{\mu\nu\alpha}$.    Here, we are mainly interested in
the real part  of the dispersion relations, so we  shall   quote the
results
for  the real  part of   $\Pi_L$,  $\Pi_T$ and $\tilde{\Pi}$ .  The
corresponding  imaginary  contributions can be obtained by analytic
continuation in the complex $k^2$ plane \cite{nos}; these will
contribute to
the damping rate in the medium and can be neglected in first
approximation.
In each of the expressions for $\Pi_L$,  $\Pi_T$ and $\tilde{\Pi}$,
that are
obtained from \Eq{pimunu2},  the integration over the time-like
component of
the momentum   and the angular integration can  be  explicitly
carried out. In
each case,  there remains a  nontrivial integral over $p_0 =
\sqrt{m^2 + |\vec
p|^2}$.  The results,   in the rest frame of the medium,  are as
follows:
\newpage
\begin{eqnarray}\label{pil}
 \Pi_L    =     \Pi_L[0]    & - & {\alpha k^2 \over  \kappa^2 }
\int_{|m|}^\infty dp_0  \bigg[  2 - \left( 4 p_0^2  + k^2 + 4 \omega
p_0
\right)  G^{-1}(p_0)
 \nonumber \\
  & + & \left( 4 p_0^2  + k^2 - 4 \omega p_0 \right)  G^{-1}(- p_0)
\bigg]
   \  \left( n_F(p_0 - \mu)  +   n_F(p_0 +  \mu)  \right)     \,,
\end{eqnarray}
\begin{eqnarray}\label{pit}
  \Pi_T   =       \Pi_T[0]   &+&
 {\alpha  \over  \kappa^2 }
\int_{|m|}^\infty dp_0  \bigg[  2 \omega^2  - \left( 4 m^2 \kappa^2
+ k^2 (
\omega + 2 p_0 )^2 \right)  G^{-1}(p_0)    \nonumber \\
  &+ &  \left(   4 m^2 \kappa^2 + k^2 ( \omega - 2 p_0)^2   \right)
G^{-1}(-
p_0) \bigg]
      \left( n_F(p_0 - \mu)  +   n_F(p_0 +  \mu)  \right)     \,,
\end{eqnarray}
\begin{equation}\label{pitil}
 \tilde{\Pi}    =    {m \alpha }  \int_{|m|}^\infty dp_0  \left[
G^{-1}(p_0) -
 G^{-1}(- p_0) \right]
 \left( \tanh{{\beta \over 2} (p_0 - \mu)}   +    \tanh{{\beta \over
2} (p_0 +
\mu)}  \right)
 \,  ,
\end{equation}
 where $\alpha = e^2/4\pi$ and  all the results refer to the real
contributions.
The auxiliary function $G(p_0)$ is defined as
\begin{equation}\label{fung}
 G(p_0)  =  \left[  \left( 2 \omega p_0 + k^2  \right)^2  -
4 \kappa^2 \left(p_0^2 - m^2  \right)\right]^{1/2}  \,  .
\end{equation}

In Eqs. (\ref{pil}) and (\ref{pit})  we have explicitly separated the
$(T= 0  ,
\mu= 0)$ vacuum polarization contribution to  $\Pi_L$ and    $\Pi_T$
. Thus,
$\Pi_L[0]$ and    $\Pi_T[0]$  represent the regularized vacuum
contribution to
these functions. In what follows we do not need to consider the
explicit
expressions for   $\Pi_L[0]$ and    $\Pi_T[0]$, as in the large $T$
or $\mu$
limit we shall here consider, the background  contributions become
dominant. On
the other hand, we have explicitly   added   the vacuum and
background
contributions for the parity violating term    because, as we  shall
see,
the two contributions to $\tilde{\Pi}$  are finite and of the  same
order.

The integrals  in   Eqs. (\ref{pil}) to (\ref{pitil}) cannot be
computed
analytically except in some limiting cases. It is instructive to
compute first
the vacuum   form of the parity  breaking term.   If we take $ T =
\mu = 0$ in
\Eq{pitil}, the integral can be easily evaluated to give
\begin{equation}\label{pitilV}
 \tilde{\Pi}_V    =
  {2 m \alpha }  \int_{|m|}^\infty dp_0  \left[ G^{-1}(p_0) -
 G^{-1}(- p_0) \right] \, = \,
 -   {m \alpha \over  k }
\ln{\left[  {  2 |m| +  k
       \over   2 |m| -  k  } \right]} \,  ,
\end{equation}
where the subindex $V$ refers to the vacuum contribution to
$\tilde{\Pi}$.
We  should remark that  this result  agrees with  the one obtained in
references  \cite{pisarski} and \cite{aitchison}, after performing a
Wick
rotation from  Euclidean to Minkowski momentum space in  the results
of those
calculations.

Let us  now turn our attention  to  the high temperature limit,
 $T \gg ( m ,  \omega , \kappa) \, ; \mu = 0$.
A systematic  high-$T$ expansion can be obtained  by expanding the
thermal distributions in Eqs.(\ref{pil}) and (\ref{pit}). However, in
this
limit
the leading  contribution  to $\Pi_T$ and $\Pi_L$  can be easily
obtained
following the ideas introduced  by  Braaten and Pisarski
\cite{braaten}
 to evaluate the hard-loop contributions  to  Feynman diagrams.
 The   hard-loop contributions    to $\Pi_T$ and $\Pi_L$  are
extracted by
 neglecting  the mass and  the external momentum  as compared with
the
internal
$p_0$ momentum in the integrals of  Eqs. (\ref{pil}) and (\ref{pit}).
Thus,
one  is left with  integrations which are elementary and lead to the
results
\begin{eqnarray}\label{piTe}
 \Pi_L(\omega,\kappa)  &= &  2  \omega^2_P   {k^2 \over \kappa^2}
\left( {\omega \over k } - 1 \right)       \, , \nonumber\\
 \nonumber\\
  \Pi_T(\omega,\kappa)   &= & 2    \omega^2_P {\omega \over \kappa^2
} \left(
\omega - k \right)
       \,,     \end{eqnarray}
in the large-$T$ limit, where for convenience we have defined  the
plasma
frequency (or plasma mass) as $  \omega^2_P   = (2 \alpha \ln{2})
T$.
We notice that the leading order  contribution to $\Pi_L$ and $\Pi_T$
is
proportional to $T$. The leading contribution arises from those
terms that
would diverge  linearly if there were no thermal distribution  to cut
the
integrations off at  $p_0 \sim {\cal O} (T)$.  It is instructive to
compare
these results with those obtained in $3 + 1$ dimensions,  where
$\Pi_T$ and
$\Pi_L$  are proportional  to $T^2$ in the high-$T$ limit
\cite{weldon}.

The  expression \Eq{pitil} for the   parity violating term  is not
divergent in
the limit $T\to \infty$, consequently the high-$T$  contribution to
$\tilde{\Pi}$  cannot be evaluated using the hard-thermal loop
method. In order
to evaluate the leading contributions to  $\tilde{\Pi}$ we    split
the
integral  in \Eq{pitil}  into two contributions:
$\int_{m}^{\infty} \, = \,  \int_{m}^{2T} \, + \,
\int_{2T}^{\infty}$.
For the first contribution,   the range of integration is $(m, 2T)$
and
  the function $\tanh{(p_0/2T)}$ can be expanded in powers of
$p_0/2T$;  the
resulting series  can be  integrated term by term and  from the
analytical
expressions obtained the   leading contribution can be computed  for
large $T$.
The  leading term  arises  from the first term of the series, which
can be
explicitly evaluated as
\begin{equation}\label{pitilT}
 \tilde{\Pi} \, = \, {\alpha \over 2}  \, {m \over T} \,  {\omega
\over k} \,
\ln{\left[  {|m| \over  T}  \right]}
    \, + \, {\cal O} \left( { m \over T} \right)
  \,  .
\end{equation}
Here,  we have assumed for simplicity that  $( m \gg  \omega \sim
\kappa)  $.
The second contribution  (coming from the integration from $2T$ to
$\infty$)
can be neglected because it is of order $m/T$.  Thus, the result
given  in
\Eq{pitilT} represents the leading contribution to the parity
breaking term at
high temperature.

{}From the previous results we can conclude that in the large-$T$
limit the
background contributions  to $\Pi_L$ and  $\Pi_T$ are dominant and
proportional
to
$T$. Remarkably,  we notice  that the   $T\to \infty$ background
contribution
to $\tilde{\Pi}$ is finite and  cancels with the vacuum contribution
\Eq{pitilV},  in such a way that  the   leading  contribution to the
total
parity violating term is proportional to $(m/T)  \ln{\left[  {|m |/
T}
\right]} $. Consequently,  in the high-$T$ limit  $\tilde{\Pi} $
vanishes asymptotically, so the parity symmetry gets restored. This
result is
similar  to the phenomena in which a   spontaneously   broken
symmetry   is
restored at high temperature  \cite{linde}. But, whereas in the
latter case the
symmetry is restored when $T$  is  above a certain critical
temperature, in the
present  situation  the  parity symmetry is only restored  as  $T \to
\infty$.
 Here we have verified the result  by calculating the leading
one-loop
contribution to the boson self-energy,  so it  is expected to be
accurate
 in the weak coupling limit,  a more detailed analysis  will be
presented
elsewhere.

Although for high temperature $\Pi_L$ and $\Pi_T$ dominate over
$\tilde{\Pi}$,
the latter is still significant if one limits oneself to calculations
that
describe phenomena of purely parity violating character.  As regards
to the
dispersion relations,  though, $\tilde{\Pi}$ can be neglected  and
\Eq{polo}
reduces to the usual  conditions   $k^2 = \Pi_L$ and   $k^2 = \Pi_T$
for the
longitudinal and transverse modes, respectively. Using the  results
given in
\Eq{piTe}  we can calculate explicitly the dispersion relations and
obtain
\begin{equation}\label{wlong}
  \omega^2    =     {\left(\kappa^2 + 2 \omega_P^2  \right)^2 \over
\left(\kappa^2 +  4 \omega_P^2  \right)  }
  \,  ,
\end{equation}
for the longitudinal oscillations  and
\begin{equation}\label{wtran}
  \omega^2   =  { \left[ \omega_P^2 + \sqrt{  \omega_P^4 + 2
\omega_P^2
\kappa^2}
\right]^2 - \kappa^4
 \over  \left( 4 \omega_P^2 - \kappa^2 \right) }
  \,  ,
\end{equation}
for the  transverse one.   We   notice that the longitudinal and
transverse
modes are characterized by the same  value of the mass:
 $\omega_P^2 = \Pi_T(\omega,   \kappa = 0) = \Pi_L(\omega,   \kappa =
0) $.
   Both modes propagate in the plasma for  frequencies above
$\omega_P$, though
no plane waves can propagate for frequencies $\omega < \omega_P$,
because
the wave number $\kappa$   becomes imaginary. The dispersion
relations in  Eqs.
(\ref{wlong}) and (\ref{wtran}) can still be used to determine
$Im[\kappa]$.
The   behavior of the space-like  excitations   can be understood if
we note
the limiting behavior:
\begin{equation}\label{pizero}
\Pi_T(\omega \to 0, \kappa) =  \tilde{\Pi} (\omega \to 0, \kappa)  =
0  \, ,
\qquad
\Pi_L(\omega \to 0, \kappa)  =  2 \omega_P^2  \equiv k_D^2  \,  ,
\end{equation}
where we have defined  the  inverse Debye length $k_D $. These
results
 show  that the well-known phenomenon  of  screening of static
electric fields
and  the fact that static magnetic fields are not screened are  still
valid  in
$2+1$ dimension  in the  limit of high temperature.

We have shown  that in this $T \gg ( m, \omega , \kappa)$ limit, the
parity
breaking
effects are negligible and the behavior of the plasma modes is
similar to that
observed in $3 + 1$ dimensions.  On the other hand, in the regime  $T
\sim m$
we expect the parity breaking term to play an  important role; in
particular
this term should modify the usual  plasma relations. However,   $T
\sim m$  is
a crossover regime which   complicates the calculations, requiring a
more
elaborate analysis which will be presented elsewhere \cite{nos}.
Fortunately,
we can consider with some more detail the degenerate limit, in which
the Fermi
energy is  much  greater than the temperature.  In this case  the
integrals in
Eqs. (\ref{pil}) to (\ref{pitil})  can be  evaluated analytically
thanks to the
fact that  the thermal  distributions can be well approximated by
step
functions. Therefore, let us consider  the  degenerate  case  where
we shall
be able to illustrate some of  the  effects  that the parity breaking
term
introduces into the plasma properties.

The exact expressions for $\Pi_T$,    $\Pi_L$  and  $\tilde{\Pi}$
for  the
degenerate plasma will be  discussed in detail   elsewhere
\cite{nos}. Here we
shall only discuss the  results in the  particular  limit $
\mu{\lower-1.2pt\vbox{\hbox{\rlap{$>$}\lower5pt\vbox{\hbox{$\sim$}}}}}
{}~\!m~\!\gg~\!\omega~\!\sim~\!\kappa$.
Before proceeding to this task, though, we  shall briefly  discuss
some of the general properties of   $\tilde{\Pi}$.

In the degenerate  limit ($\mu  \gg T  \approx 0$)  the integral  in
\Eq{pitil}
can be calculated   analytically leading to the result
\begin{equation}\label{pitilMu}
 \tilde{\Pi}  =    {2 m \alpha }  \int_\mu^\infty dp_0  \left[
G^{-1}(p_0) -
 			G^{-1}(- p_0) \right]     \, = \,  -  {m
\alpha \over   k }
			\ln{\left[  { \left(2 \mu   + \omega \right)
k +  G(\mu) \over
			 \left( 2 \mu    -  \omega \right) k  +  G(-
\mu)   }  \right]}
 \,  ,
\end{equation}
where $G$ is the function defined in \Eq{fung}  and we have assumed
$\mu > 0$.
It is interesting to notice that the expression for  $\tilde{\Pi}$
has exactly
the same structure as the vacuum contribution (see \Eq{pitilV}), but
now the
range of integration for the fermion energy is $ \mu \leq p_0 \leq
\infty$.
This may look strange because the   fermions  of the background
fill up the
levels with energies  in the region   $ m \leq  p_0  \leq\mu$, but
what happens
is   that  the background  contribution has exactly the same
structure as the
vacuum one, but with opposite sign.  Consequently  the background
contribution
$(\int_{m}^{\mu})$ cancels part of  the vacuum contribution
$(\int_{m}^{\infty})$   so one is left with the result given in
\Eq{pitilMu}.

It is worth noticing that in high density regime,  $\mu \gg
(m,\omega,\kappa)$,
 the parity breaking term  \Eq{pitilMu} is approximately
$  \tilde{\Pi} = - \alpha (\omega/k)(m/\mu)$.
Thus,   the parity symmetry  is also restored in the high density
limit.
Again, it is interesting  to compare this result with the one
obtained  by
Harrington and Yildiz \cite{fulanito},   who found that the
spontaneous
symmetry breaking is  restored,  not only in the high-$T$ limit, but
also  in
the high density limit.  In this high-density limit  the propagating
modes
reduce again  to longitudinal  and  transverse   ones,  with the
dispersion
relations given in Eqs. (\ref{wlong})  and (\ref{wtran}),
respectively, but
with  the plasma frequency  now given by  $\omega^2_P =  \alpha \mu$,
$i.e.$
the results in the high density limit are equivalent to those in the
high-$T$
limit,  with $(\ln{2})  T$ merely replaced by  $ \mu/2$.

In order to illustrate the effect that the parity violation term can
induce in
the plasma   relations  we consider a particular  limit  in which
$\mu$  and
$m$ are very close to  each other but still much  larger than
$\omega$ and
$\kappa$. If we define the Fermi momentum $\kappa_F$ according to the
relation
$\mu = \sqrt{\kappa_F^2 + m^2}$
we can write these conditions as $ \mu {\
\lower-1.2pt\vbox{\hbox{\rlap{$>$}\lower5pt\vbox{\hbox{$\sim$}}}}\ }
m
\gg  (\omega , \kappa)$ and $\mu^2 \gg \kappa_ F^2$. These
observations allow
us to approximate the exact expression \Eq{pitilMu} and the
corresponding ones
for  $ \Pi_L$, $\Pi_T$ to obtain the polarization tensor in this
limit.
For $\omega \to  0 $ we find that the results given in \Eq{pizero}
still hold.
Notice,  that according to this result  there is not magnetic
screening.
The inclusion of an explicit Chern-Simons term   produces magnetic
screening
\cite{deser}; however here $\tilde{\Pi}$ vanishes at $\omega = 0$.
For  any
value of
$\kappa$  such that  $\kappa  \kappa_F \ll \omega \mu $,  we obtain

\begin{eqnarray}\label{piden}
  \Pi_L  &= &   \alpha {\kappa_F^2 \over  \mu}   -  \alpha {
\kappa_F^4
\over 2 \mu^3 } \left(  {\kappa \over \omega } \right)^2
   \, , \nonumber\\
 \nonumber\\
  \Pi_T  &= &    \alpha {\kappa_F^2 \over  \mu}   +   \alpha {
\kappa_F^2
\left(  \mu^2 + m^2  \right)
\over 2 \mu^3 } \left(  {\kappa \over \omega } \right)^2
  \, , \nonumber\\
 \nonumber\\
  \tilde{\Pi} & = &  -  \alpha {m \over  \mu }
\left[ 1 - {1\over 2} \left( {\kappa \over \omega } \right)^2
 \left( {\kappa_F \over \mu } \right)^2 \right]
    \,.     \end{eqnarray}

The dispersion relations are deduced  by solving the pole equation
(\ref{polo}).
 In  the region in which the plane waves can propagate, the relation
 $\kappa  \kappa_F \ll \omega \mu $ holds and  Eqs. (\ref{piden})
apply.
 \Eq{polo} then gives rise to two independent modes with different
masses: the
$(+)$ mode given by
\newpage
\begin{eqnarray}\label{omega+}
 \omega^2  &= &
 \omega_{+}^2 + \left(1 + {\kappa_F^2 \over \alpha \mu}   \right)
\kappa^2
\qquad \qquad  \qquad \qquad   \qquad
when \qquad \kappa \ll \omega_{+}   \, , \nonumber\\
 \nonumber\\
 \omega^2  &= & \kappa^2 + \left( {\alpha m \over \mu }  \right)^2
\left( 1 + {\kappa_F^2 \mu \over \alpha m^2}
\left[ {5\over 2} + {m^2 \over 2 \mu^2} \right]\right)  \qquad when
\qquad
\kappa \gg \omega_{+}
       \,,     \end{eqnarray}
where we have defined the plasma frequency
$\omega_+^2 = (\alpha^2 m^2 / \mu^2) \left( 1 + (2 \mu \kappa_F^2/
\alpha m^2 )
 \right)$.
This mode corresponds to a right-handed longielliptic wave;  indeed,
assuming that the wave propagates along the $x-$axis,
 the polarization for small $\kappa$  is  approximately given  as \\
 $\vec{E}  \propto (1, i(m/|m|)[ 1  + \kappa^2 \mu^2/ (2\alpha^2
m^2)])$.
For the $(-)$ mode we find
\begin{eqnarray}\label{omega-}
 \omega^2  &= &
 \omega_{-}^2 + \left(1 + {\kappa_F^4 \over \mu^2 \left( \omega_{-}^2
+
\kappa^2 \right) } \right)  \kappa^2
\qquad\qquad \qquad   \qquad \,\,\,
 when \qquad \kappa \ll \omega_{-}   \, , \nonumber\\
 \nonumber\\
 \omega^2  &= & \kappa^2 + {\kappa_F^8 \left( m^2 + 5 \mu^2 \right)^2
\over
24 m^2  \mu^8}
\qquad\qquad \qquad   \qquad\qquad \qquad
when  \qquad \kappa \gg \omega_{-}
       \,,     \end{eqnarray}
with $\omega_{-}^2 = (\kappa_F^4/m^2) \left(1 - (2\mu
\kappa_F^2/\alpha m^2)
\right)$.
This is a left-handed mode; $e.g.$  the polarization  can be
approximately
written as
$\vec{E}  \propto (1,  -i(m/|m|)[ 1 + \kappa^2 m^2/(2 \kappa_F^4)
])$.
Notice that  handedness of the modes will be reversed if the sign of
$m$ is
changed.

We have demonstrated that  both in the high-temperature
and the high-density limits the parity symmetry is restored in
$QED_3$. In this
case the plasma properties are similar to those of the $3 +1 $
dimensional
system.
However in general,  the propagating modes  of the photon  consists
of two
longielliptic
waves with different values for the masses.  The appearance
of two different  masses  for the gauge fields in  $2 + 1$ parity
breaking
systems is known for models that include  the Higgs mechanism and an
explicit
Chern-Simons term.
It was shown by Pisarski and Rao \cite{pisarski2} that  the combined
effect of
the
Chern-Simons term with the mass induced by the spontaneous symmetry
breaking
gives rise to  two  modes with different masses.
Here we have shown a similar phenomenon in which the combined effect
of the
plasma with the radiatively induced  parity breaking term  results
into two
independent plasma modes, longielliptically polarized and with two
different
masses.


\newpage


\begin{thebibliography}{99}

\bibitem{jackiwplano} For a review see
         R. Jackiw, in  ``Physics geometry and Topology'', ed. by H.
C. Lee
         NATO ASI Series B: Physics Vol. 238,  (Plenum Press, New
York , 1990).

\bibitem{hall} R. B. Laughlin, \prl{60}{2677}{1988};  A. Fetter, C.
Hanna and
 R. B. Laughlin, Phys. Rev. B{\bf39} (1989) 9679; S. Girving and R.
Prange,
The Quantum Hall Effect (Spinger -Verlag, Berlin 1987).

\bibitem{deser}   S. Deser, R. Jackiw and S. Templeton,
\prl{48}{975}{1982}; S.
Deser, R. Jackiw and S. Templeton, \annphys{140}{372}{1982}; N.
Redlich,
\prl{52}{18}{1984}.

\bibitem{pisarski}    R. D. Pisarski, \prd{29}{2423}{1984}.

\bibitem{linde} A. D. Kirzhnits and A. D. Linde, \plb{42}{471}{1972};
 S. Weinberg, \prd{9}{3357}{1974}; L. Dolan and R. Jackiw,
\prd{9}{3320}{1974}.

\bibitem{chernT} K. S. Babu, A. Das and P. Panigrahi,
\prd{36}{3725}{1987};
E. R. Poppitz, \plb{252}{417}{1990};
L. Moriconi, \prd{44}{R2950}{1991};
M. Burgess, \prd{44}{2552}{1991}.

\bibitem{aitchison}  I. J. R. Aitchison, C. D. Fosco and J. Zuk,
\prd{48}{5895}{1993}.

\bibitem{review}  For a review see   N. P. Landsman  and Ch. G. van
Weert,    \physrep{145}{141}{1987}.

\bibitem{kobes} R. Kobes,  \prl{67}{1384}{1991}.

\bibitem{niemi}   A. J. Niemi and G. W. Semenoff,
\annphys{152}{105}{1984};
                              R. L. Kobes and G. W. Semenoff,
\npb{260}{714}{1985}.

\bibitem{nieves}   J. F. Nieves, \prd{42}{4123}{1990}.


\bibitem{weldon}    H. A. Weldon, \prd{26}{1394}{1982}.

\bibitem{nos} M. Klein-Kreisler  and M.  Torres, to be published.

\bibitem{braaten}  E. Braaten and R. D. Pisarski,
\npb{337}{569}{1990};
\ibid{339}{310}{1990};
                  \prd{45}{R1827}{1992}; see also. H. A. Weldon,
\prd{26}{2789}{1982}.

\bibitem{fulanito} B. J Harrington and A. Yildiz,
\prl{33}{324}{1974}.

\bibitem{pisarski2} R. Pisarski and S. Rao, \prd{32}{2081}{1985}.

\end{thebibliography}
\end{document}